\documentstyle[prl,aps]{revtex}        
\begin{document}
\draft
\twocolumn
\title{On the failure of Bell's theorem} 
\author{Gyula Bene}
\address{Institute for Solid State Physics, E\"otv\"os University,
M\'uzeum krt. 6-8, H-1088 Budapest, Hungary}
\date{\today}
\maketitle

\begin{abstract}
Using a new approach to quantum mechanics \cite{Bene1}, \cite{Bene2}
we revisit Hardy's proof \cite{Hardy} for Bell's theorem 
and point out a loophole in it. We also demonstrate on this example that quantum
mechanics is a local realistic theory.
\end{abstract}
\pacs{03.65.Bz}
\narrowtext

Bell's famous theorem\cite{Bell} states that no local realistic theory can fully
reproduce quantum mechanical correlations. There are a number of 
very clear and convincing proofs for this statement, both using inequalities
\cite{Bell}-\cite{bell3} and without them \cite{other1}-\cite{other4}, \cite{Hardy}. 
Certainly, all these proofs are
logically correct, so they can fail only if one can show that tacitly 
all the time there has been
made some natural-looking extra assumption beyond 
realism and locality. If such an assumption is uncovered, then
the usual contradiction appearing in the proofs of Bell's theorem 
may imply the failure of this extra assumption rather than that of the principle of
locality or realism.
It has recently been demonstrated on the example of Bell's inequality 
that it is actually the case \cite{Bene1}, \cite{Bene2}. In the present paper
we point out the failure of Bell's theorem by revisiting Hardy's proof \cite{Hardy}
which is probably the simplest and most powerful one and does not use inequalities.

Suppose that an EPR-Bell
experiment is performed on a two-particle system. The results of the
measurements (in a particular run) are denoted by $a$ and $b$, respectively,
and the hidden variable characterizing the original two-particle state
be $\lambda$. Realism is equivalent with the statement that each of
 these three quantities is an element of the reality. 
 The extra assumption mentioned above is the following: 
 
 {\em realism (as expressed above) implies that the quantities
$a$, $b$, $\lambda$ can be in principle compared.}

Without this assumption one could not assign a joint probability to the set
$\{a,\,b,\,\lambda\}$ or could not
speak about events when these quantities simultaneously have certain 
definite values. Certainly this assumption looks rather obvious, 
however, it does not follow on a purely logical ground. 
Formally, the comparability of the above quantities would mean
that also the {\em set} $\{a,\,b,\,\lambda\}$ is an {\em element} of reality.
However, the condition that $a$, $b$ and $\lambda$  are elements of the reality
implies only that the set $\{a,\,b,\,\lambda\}$ 
is a {\em subset} of the reality. 

In the recent theory\cite{Bene1}, \cite{Bene2} which underlies our
considerations, the role of the quantities $a$, $b$ and $\lambda$
is played by certain quantum states, and their reality means that 
any of them can be determined by a suitable measurement with unit probability,
without changing that state. It does not imply, however, that all the three
appropriate measurements can be performed {\em simultaneously}
without changing the states and their correlations. 
Indeed, the measurement which does not disturbe the state
corresponding to $\lambda$ does change another state, which leads to a
change of the correlation between $a$ and $b$. Therefore, the states
corresponding to $a$, $b$ and $\lambda$ exist, yet
it is physically meaningless to speak about events of the type $a\wedge b\wedge \lambda$.

Let us briefly review the theory \cite{Bene1}, \cite{Bene2}. According to it
one and the same system $S$ can be characterized by a multitude of states $\hat \rho_S(R_1),\,
\hat \rho_S(R_2),\, ...$, where the {\em quantum reference systems} $R_1,\, R_2,\, ...$
are also physical systems which contain the system $S$. These states are all elements of the
reality, as in principle any of them can be learned with unit probability by a suitable measurement 
which does not disturbe that state. Mathematically these states are positive definite
hermitian operators with unit trace, and they act on the Hilbert space of the system $S$.
The theory keeps the Schr\"odinger equation unchanged, but 
rejects the idea of the collapse of the wave function. The interrelation of the
states is specified by a few new postulates which allow for 
the description of the measurements themselves as usual interactions
between physical systems. The postulates (which will be needed in the 
discussion below) are the following:

{\bf Postulate 1. \em The state $\hat \rho_S(S)$  is always a single
dyad $|\psi_S><\psi_S|$.} 
We shall call both $\hat \rho_S(S)$ and $|\psi_S>$
the internal state of $S$.

{\bf Postulate 2.} $\hat \rho_S(R)=Tr_{R\setminus S} |\psi_R><\psi_R|$
 
{\bf Postulate 3.\em  If the reference system $R=I$ is an isolated one
\footnote{We define an isolated system as
such a system which has not been interacting 
with the outside world. In contrast, a closed system
is such a system that is not interacting with any other
system at the given instant of time 
(but might have interacted in the past).}
then the state $\hat \rho_S(I)$ commutes with the
internal state $\hat \rho_S(S)$.}

{\bf Postulate 4.\em  The result of a measurement is contained 
unambigously in the internal state of the measuring device.}

{\bf Postulate 5. \em If there are $n$ ($n=1,\;2,\;3,\;...$) disjointed physical systems, 
denoted by
\hfill\break
$S_1, S_2, ... S_n$, all contained in the isolated reference 
system $I$ and 
having the 
possible internal states
$|\phi_{S_1,j}>,...,|\phi_{S_n,j}>$, respectively, 
then the joint
probability that $|\phi_{S_i,j_i}>$ 
coincides with the internal state of $S_i$ ($i=1,..n$)
is given by
\begin{eqnarray}
P(S_1,j_1,...,S_n,j_n)\mbox{\hspace{3cm}}\nonumber\\
=Tr_{S_1+...+S_n} [\hat \pi_{S_1,j_1} 
...\hat \pi_{S_n,j_n}\hat \rho_{S_1+...+S_n}(I)],\label{u5}
\end{eqnarray}
where $\hat \pi_{S_i,j_i}=|\phi_{S_i,j_i}><\phi_{S_i,j_i}|$.}
\vskip0.5cm
Let us recall now Hardy's proof \cite{Hardy} for Bell's theorem\footnote{Nearly the same notations
are used as in \cite{Hardy}, however, we renormalize the states by some $-1$-s and $i$-s,
in order to have everywhere real coefficients.}. 
The initial state of the two particle system $P_1+P_2$ is 
\begin{eqnarray}
|\psi>_{P_1+P_2}=\alpha |+_1> |+_2> - \beta |-_1> |-_2>\,,\label{e1}
\end{eqnarray}
where $\alpha$ and $\beta$ are real positive numbers with 
$\alpha^2+\beta^2=1$ and $\alpha \ne \beta$. 
One defines the normalized states
\begin{eqnarray}
|u_i>=b|+_i>+a|-_i>\nonumber\\
|v_i>=-a|+_i>+b|-_i>\,,\nonumber
\end{eqnarray}
where
\begin{eqnarray}
a=\sqrt{\frac{\alpha}{\alpha+\beta}}\,,\quad b=\sqrt{\frac{\beta}{\alpha+\beta}}\,.\nonumber
\end{eqnarray}
Obviously, $<u_i|v_i>=0$. Let us define further normalized states
\begin{eqnarray}
|c_i>=A|u_i>+B|v_i>\nonumber\\
|d_i>=-B|u_i>+A|v_i>\,,\nonumber
\end{eqnarray}
where
\begin{eqnarray}
A=\sqrt{\frac{\alpha \beta}{1-\alpha\beta}}\,,\quad B=\frac{\beta-\alpha}{\sqrt{1-\alpha\beta}}\,.\nonumber
\end{eqnarray}
Again, $<c_i|d_i>=0$ follows. Consider now the observables $U_i$ and $D_i$ 
corresponding to the operators $\hat U_i=|u_i><u_i|$ and $\hat D_i=|d_i><d_i|$, respectively.
Now one considers four different measurements and calculates their outcome according to standard quantum mechanics.
\begin{enumerate}
\item If $U_1$ and $U_2$ are measured (where the indices refer to the corresponding particles),
then $U_1 U_2=0$.

\item If $D_1$ and $U_2$ are measured, then $D_1=1$ implies $U_2=1$.

\item If $U_1$ and $D_2$ are measured, then $D_2=1$ implies $U_1=1$.

\item If $D_1$ and $D_2$ are measured, then $D_1=1$ and \hfill\break $D_2=1$ happens with probability\hfill\break
$\alpha^2\beta^2(\alpha-\beta)^2/(1-\alpha\beta)^2$.
\end{enumerate}

Now the argument goes on as follows. Suppose that one measures $D_1$ and $D_2$,
and at some (unknown) value $\lambda$ of the hidden parameter (which characterizes the
original two particle state) $D_1(\lambda)=1$ and $D_2(\lambda)=1$ occurs. As can be seen above (cf. item 4.),
there is a finite probability for that. But then item 2. and item 3. 
imply (provided that the principle of locality holds), that at the same value of the 
hidden parameter $U_2(\lambda)=1$ and $U_1(\lambda)=1$, 
that contradicts item 1., which states that
this situation cannot happen. The conclusion of Ref.\cite{Hardy} is that 
the principle of locality fails.

Let us consider now the above situation from the point of view of the new theory \cite{Bene1}, \cite{Bene2}.
The initial state $|\psi>_{P_1+P_2}$ of the two particle system
corresponds to the initial internal state $\hat \rho_{P_1+P_2}(P_1+P_2)$ (cf. {\bf Postulate 1.}). 
Let us calculate now the state $\hat \rho_{P_1}(P_1+P_2)$. According to
{\bf Postulate 2.}, it is
\begin{eqnarray}
\hat \rho_{P_1}(P_1+P_2)=|+_1> \alpha^2 <+_1|+|-_1> \beta^2 <-_1|\,.
\end{eqnarray}
Assuming that initially $P_1+P_2$ is an isolated system\footnote{In a more rigorous treatment one 
should also include the preparation process of the state $|\psi>_{P_1+P_2}$. This, however, does not
have any significant influence on the final results. For such a treatment (in connection
with the EPR paradox) see Ref. \cite{Bene1}.} we can apply {\bf Postulate 3.}. We get that
the internal state $\hat \rho_{P_1}(P_1)$ is either $|+_1> <+_1|$ with probability $\alpha^2$ (as implied by
{\bf Postulate 5.} with $n=1$) or \hfill\break $|-_1 ><-_1|$ with probability $\beta^2$.
Similarly, we get that $\hat \rho_{P_2}(P_2)$ is either $|+_2> <+_2|$ with probability $\alpha^2$ 
 or $|-_2> <-_2|$ with probability $\beta^2$. Applying now {\bf Postulate 5.} with $n=2$ we get
 that  $\hat \rho_{P_1}(P_1)=|+_1 ><+_1|$ and $\hat \rho_{P_2}(P_2)=|-_2 ><-_2|$ cannot simultaneously happen,
 and $\hat \rho_{P_1}(P_1)=|-_1 ><-_1|$ and $\hat \rho_{P_2}(P_2)=|+_2> <+_2|$ cannot happen, either.
 In other terms, if the internal state of $P_1$ is $|+_1>$, then the internal state of $P_2$ is $|+_2>$,
and if the internal state of $P_1$ is $|-_1>$, then the internal state of $P_2$ is $|-_2>$. We shall
say briefly that the internal state of $P_1$ and that of $P_2$ are uniquely related.

Let us denote the eigenstates of the measured observable by $|\xi^{(i)}_j>$ ($i,j=1,2$).
This means that $|\xi^{(i)}_1>$ ($|\xi^{(i)}_2>$) can be either $|u_i>$ ($|v_i>$) or
$|c_i>$ ($|d_i>$). The dynamics of the measurement be given by the relations (these
are assumed to be approximately the same as that one would get from 
the time dependent Schr\"odinger equation)
\begin{eqnarray}
|\xi^{(i)}_j>|m^{(i)}_0>\,\rightarrow \,|\xi^{(i)}_j>|m^{(i)}_j>\,,\label{dyn}
\end{eqnarray}
where $|m^{(i)}_0>$ is the initial internal state of the $i$-th measuring
device and $|m^{(i)}_j>$ describes the measuring device
when it displays the $j$-th result. 

The initial internal state of the system $P_1+P_2+M_1+M_2$ is
\begin{eqnarray}
\left(\alpha |+_1 >|+_2 >- \beta |-_1> |-_2>\right)|m^{(1)}_0>|m^{(2)}_0>\,.
\end{eqnarray}
Using again {\bf Postulate 1.}, {\bf Postulate 2.} and {\bf Postulate 3.} as above, one
can show that the initial internal state of the subsystem $P_1+M_1$ is
either $|+_1>|m^{(1)}_0>$ or \hfill\break $|-_1>|m^{(1)}_0>$, so the initial internal state of $P_i$
is uniquely related to that of $P_i+M_i$. Moreover, as $P_i+M_i$ is a closed system ,
its initial internal state is uniquely related to its final internal state. This latter is
(using Eq. (\ref{dyn})) either
\begin{eqnarray}
|\phi_+>=\sum_j <\xi^{(i)}_j|+_1>|\xi^{(i)}_j>|m^{(i)}_j>\label{sv1}
\end{eqnarray}
or
\begin{eqnarray}
|\phi_->=\sum_j <\xi^{(i)}_j|-_1>|\xi^{(i)}_j>|m^{(i)}_j>\,.\label{sv2}
\end{eqnarray}
Thus we conclude that 
the initial internal state of $P_i$ is uniquely related to the final internal state of $P_i+M_i$.
We can check this if we calculate first the final internal state of the isolated system
$P_1+P_2+M_1+M_2$ and from that determine what can be the final internal state of $P_i+M_i$.
We get for the final internal state of $P_1+P_2+M_1+M_2$
\begin{eqnarray}
\sum_{j,k} \left(\alpha<\xi^{(1)}_j|+_1><\xi^{(2)}_k|+_2>\right.\nonumber\\\left.
-\beta<\xi^{(1)}_j|-_1><\xi^{(2)}_k|-_2>\right)\nonumber\\
\times|\xi^{(1)}_j>|m^{(1)}_j>|\xi^{(2)}_k>|m^{(2)}_k>\,.\label{final}
\end{eqnarray}
Using {\bf Postulate 2.} we get for\hfill\break $\hat \rho_{P_1+M_1}(P_1+P_2+M_1+M_2)$
\begin{eqnarray}
\sum_{j,j',k} \left(\alpha<\xi^{(1)}_j|+_1><\xi^{(2)}_k|+_2>\right.\nonumber\\\left.
-\beta<\xi^{(1)}_j|-_1><\xi^{(2)}_k|-_2>\right)
\nonumber\\
\times\left(\alpha<\xi^{(1)}_{j'}|+_1><\xi^{(2)}_k|+_2>\right.\nonumber\\\left.
-\beta<\xi^{(1)}_{j'}|-_1><\xi^{(2)}_k|-_2>\right)^*\nonumber\\
\times|\xi^{(1)}_j>|m^{(1)}_j><m^{(1)}_{j'}|<\xi^{(1)}_{j'}|\\
=|\phi_+>\alpha^2<\phi_+|+
|\phi_->\beta^2<\phi_-|\,.\nonumber
\end{eqnarray}
Here the completeness relation $\sum_k |\xi^{(2)}_k><\xi^{(2)}_k|=\hat 1$ has been used.
As we see, the eigenstates of\hfill\break  $\hat \rho_{P_1+M_1}(P_1+P_2+M_1+M_2)$
are just (\ref{sv1}) and (\ref{sv2}). Therefore, according to {\bf Postulate 3.} we find that
the internal state of $P_1+M_1$ is indeed either (\ref{sv1}) or (\ref{sv2}), moreover,
{\bf Postulate 5.} implies that the corresponding probability is $\alpha^2$ and $\beta^2$,
respectively. 

If we want to know what the result of a measurement is, according to {\bf Postulate 4.}
we have to calculate the internal state of the corresponding measuring device. In order to do that
first we determine e.g.\hfill\break  $\hat \rho_{M_1}(P_1+P_2+M_1+M_2)$ from Eq. (\ref{final}) by
applying {\bf Postulate 2.}. We get
\begin{eqnarray}
\hat \rho_{M_1}(P_1+P_2+M_1+M_2)
=\sum_j |m^{(1)}_j>p_j<m^{(1)}_j|\,,
\end{eqnarray}
where $p_j=\alpha^2\left|<\xi^{(1)}_j|+_1>\right|^2
+\beta^2\left|<\xi^{(1)}_j|-_1>\right|^2$.
Let us emphasize that it is independent of the second measurement process.
According to {\bf Postulate 3.} this implies that the internal state of the first measuring device
is either $|m^{(1)}_1>$ or $|m^{(1)}_2>$. This is what we expected, as these states describe
the measuring device displaying a definite result. The corresponding probabilities
(according to {\bf Postulate 5.}) are the eigenvalues $\alpha^2\left|<\xi^{(1)}_j|+_1>\right|^2
+\beta^2\left|<\xi^{(1)}_j|-_1>\right|^2$. Note that this expression may be interpreted
as one intuitively expects: $\alpha^2$ and $\beta^2$ are the probabilities that
the initial internal state of $P_1$ is $|+_1>$ or $|-_1>$, respectively, while
the factors $\left|<\xi^{(1)}_j|+_1>\right|^2$, $\left|<\xi^{(1)}_j|-_1>\right|^2$
are the conditional probability to get the $j$-th result provided that the
the initial internal state of $P_1$ has been $|+_1>$ or $|-_1>$, respectively.

Up to now we have seen that the initial internal state of $P_1$ uniquely determines 
the first measurement process (i.e., the time evolution of the internal state of $P_1+M_1$)
and determines the result of the first measurement in the usual probabilistic sense.
The analogous statement holds for the second measurement. We have also seen that the
two measurement processes do not influence each other. Therefore, our theory satisfies
the principle of locality, and the role of $\lambda$ (the hidden variable) is 
played by the initial internal state of one of the particles 
\footnote{It is enough to consider only one of them, as
their initial internal states are uniquely related.}. 

Let us consider now the correlations between the measurements. As we already know
the possible internal states of the measuring devices,
{\bf Postulate 5.} (with $n=2$) can be directly applied.
The result is
\begin{eqnarray}
P(M_1,j,M_2,k)=\left|\alpha<\xi^{(1)}_j|+_1><\xi^{(2)}_k|+_2>\right.\nonumber\\
\left.
-\beta<\xi^{(1)}_j|-_1><\xi^{(2)}_k|-_2>\right|^2\,.
\end{eqnarray}
Inserting here $|u_i>$ ($|v_i>$) or $|c_i>$ ($|d_i>$) for $|\xi^{(i)}_1>$ ($|\xi^{(i)}_2>$) we can
readily recover the above items 1.-4., i.e., the standard quantum mechanical predictions. 
This means that our theory is a counterexample to Bell's theorem, as it is a local realistic theory,
yet it fully reproduces the standard quantum mechanical correlations. 

What is then wrong with the argument of \cite{Hardy}? That argument relies on the existence of an event when
$D_1=1$, $D_2=1$ {\em and} $\lambda$ has a definite value. In our case this can be translated to
the statement that there is a nonzero probability that $\{$ \em initially 
the internal state of $P_1$ coincides with  e.g.
 $|+_1>$ {\bf and} finally the internal state of $M_1$  coincides with $|m^{(1)}_2>\quad $ 
{\bf and} the internal state of $M_2$  coincides with $|m^{(2)}_2>\,\}$.\rm
However, it turns out that such a probability cannot be
defined within the framework of our theory, and it is physically meaningless to speak about the above event.
Let us show this in more detail. The first difficulty when trying to find a suitable
expression for the probability in question arises because we want to compare
states given at different times, while the theory provides us only with 
equal time correlations. However, the initial internal state of $P_1$ is uniquely related
to the final internal state of $P_1+M_1$, so we may try to find the probability of the event that
$\{$ \em  
the final internal state of $P_1+M_1$ coincides with  
 $\sum_j <\xi^{(1)}_j|+_1>|\xi^{(1)}_j>|m^{(1)}_j>$\footnote{In this case $|\xi^{(1)}_1>=|c_1>$ and
 $|\xi^{(1)}_2>=|d_1>$.} 
 {\bf and} the final internal state of $M_1$  coincides with $|m^{(1)}_2>\quad $ 
{\bf and} the final internal state of $M_2$  coincides with $|m^{(2)}_2>\,\}$.\rm
But now the problem is that the systems involved (i.e., $P_1+M_1$, $M_1$ and $M_2$) are not
disjointed, while {\bf Postulate 5.} applies for disjointed systems. If we yet try to apply
Eq. (\ref{u5}) directly,
we get the expression 
\begin{eqnarray}\footnotesize
\alpha <+_1|\xi^{(1)}_2><+_2|\xi^{(2)}_2>\nonumber\\\times
\left(\alpha <\xi^{(1)}_2|+_1><\xi^{(2)}_2|+_2>\right.\nonumber\\
\left.-\beta <\xi^{(1)}_2|-_1><\xi^{(2)}_2|-_2>\right)\\
=\frac{\alpha^2 \beta^4 (\beta-\alpha)}{(\alpha+\beta)(1-\alpha\beta)^2}\,.\nonumber
\end{eqnarray}
If $\alpha > \beta$, this expression becomes negative. If $\alpha < \beta$, this expression
becomes larger than $P(M_1,2,M_2,2)=P(D_1,D_2)=\alpha^2\beta^2(\alpha-\beta)^2/(1-\alpha\beta)^2$,
showing that in that case the "probability" of the event obtained by replacing $|+_1>$ with $|-_1>$ 
is negative. Thus we see that no reasonable definition of the probability in question can be given.
This also means that the events $(a,b)$ cannot be classified according to the value of $\lambda$,
hence no contradiction can be derived.

In more physical terms the situation is the following. As it has been shown in Refs. \cite{Bene1}, \cite{Bene2}
any state can in principle be measured without disturbing it.  However, if we try to 
perform such nondisturbing measurements on several systems simultaneously, it
will not change the states and their correlations only if the systems are disjointed. In that case
the measurement of one system does not disturbe the other one. But if
the systems are not disjointed, a measurement which does not disturbe one particular
state will disturbe another one. Therefore, in such a case one should conclude, that
although the states individually exist, they cannot be compared.\footnote{If we try
to compare them via extra measurements, these measurements will change the correlation. 
E.g. if we perform a nondisturbing 
measurement on $P_1$ still before the other two measurements in order to record
the initial internal state of $P_1$, then the above joint probability
becomes meaningful, but due to the measurement $P(a,b)=P(M_1,j,M_2,k)$ changes \cite{Bene2}
so much that no contradiction can be constructed.}
 This circumstance has not been
taken into account in the proofs of Bell's theorem. This made possible the construction
of the local realistic theory \cite{Bene1}, \cite{Bene2}, which is an explicit counterexample to Bell's theorem.

\centerline{\bf Acknowledgements}

 The author wants to thank for the hospitality of the {\em Institut f\"ur 
 Festk\"orper\-physik, Forschungszentrum J\"ulich GmbH} where a 
 substantial 
 part of the work has been done. 
 This work has been partially supported by the OMFB (Hungary) and the BMBF (Germany)
 within the framework of a Hungarian-German joint project, the Hungarian Academy of 
 Sciences
 under Grant Nos. OTKA T 017493, OTKA F 17166 and OTKA F 19266.

\end{document}